\documentclass[9pt,twocolumn,twoside]{osajnl}
\usepackage{graphicx}
\journal{ol} % Choose journal (ao, aop, josaa, josab, ol)
\usepackage{adjustbox}
\usepackage{amsmath}
\setboolean{shortarticle}{true} 
\title{Non-invasive Focusing Through Scattering Layers Using Speckle-Correlations}

\author[1]{Galya Stern}
\author[1 *]{Ori Katz}

\affil[1]{Department of Applied Physics, The Rachel and Selim Benin School of Computer Science and Engineering, The Hebrew University of Jerusalem}

\affil[*]{Corresponding author: orik@mail.huji.ac.il}

\ociscodes{(290.0290) Scattering; (030.6140) Speckle}

\begin{abstract}
Angular speckle correlations known as the 'memory-effect' have recently been exploited for non-invasive imaging through scattering layers. Here, we show how the imaging information obtained from speckle correlations can be used as a noninvasive feedback mechanism for wavefront shaping. We utilize this feedback to demonstrate guide-star free noninvasive focusing of coherent light through highly scattering layers.

\end{abstract}

\setboolean{displaycopyright}{false}

\begin{document}

\maketitle

Focusing light through complex scattering samples, such as biological tissue, is a long sought-after capability, with great importance in many applications such as deep tissue microscopy, and non line-of-sight imaging. Such optical focusing is challenging since the examined tissue or complex samples are highly scattering due to the spatial inhomogeneities in their refractive index. Such multiple scattering causes any conventionally focused beam to diffuse, preventing conventional high-resolution optical microscopy \cite{review1}. To overcome scattering, deep-tissue optical imaging techniques either combine light and sound, or computationally tackle the  diffuse optical tomography (DOT) inverse-problem \cite{review1}. Unfortunately, these approaches suffer from a resolution at increased penetration depths that is significantly below the optical diffraction-limit \cite{review1}. 

Recently,  wavefront-shaping \cite{mosk2,mosk1} has emerged as a solution for refocusing light through thick scattering samples. In wavefront-shaping, a computer-controlled spatial light modulator (SLM) is used to tailor the input optical wavefront, such that it interferometrically inverses the effect of scattering, and focuses the beam to a diffraction-limited spot at a desired target point. However, all wavefront-shaping techniques to date depend upon a feedback from a  'guide-star' at the target \cite{horstmeyer2015guidestar}, requiring either special labeling or a complicated acousto-optical measurement.
Here, we propose an all-optical, guide-star free, wavefront-shaping approach that is based on speckle-correlations, allowing non-invasive focusing. Our approach estimates the spatial field distribution of coherent light  \textit{inside} a complex sample from an image of the scattered light recorded \textit{outside} the sample. It then utilizes the estimated pattern of the field inside the sample as a feedback for wavefront-shaping.  

When a scattering sample is illuminated by coherent light, interference of the scattered waves gives rise to random speckle patterns \cite{goodman}. While speckle patterns are usually considered as a hurdle for imaging, their complex structure contains information on the incident wave. In particular, inherent speckle correlations persist even deep beyond the transport mean-free-path in scattering samples \cite{memory1,memory2Feng,Review2,Judkewitz}. Such 'memory-effect' angular correlations were recently exploited for non-invasive diffraction-limited imaging through scattering layers, either by scanning a speckled beam on the target \cite{bertolotti}, or by a single-shot measurement of the scattered light pattern \cite{incoherent_imaging}. In the latter, the  intensity autocorrelation of spatially-incoherent light inside the scattering medium is estimated from the scattered light distribution outside the medium. Here, we extend this approach to spatially-coherent illumination, and utilize the estimated field autocorrelation inside the sample as a feedback for wavefront shaping.

 %\textbf{ADD HERE A REFERENCE TO https://www.nature.com/articles/nphoton.2012.150}
%\textcolor{red}{was always there}

%Different methods have been proposed to overcome scattering in imaging, each with its own setback and strengths. In spectral analysis and phase conjugation,   the scatterer matrix is measured using a known incident light, and is used in calibration to decode the image of an object. Ghost imaging and holographic processes use a reference beam to be compared with the coincidence counts in order to receive information about the object. In-line holography avoids the need of a reference beam by taking the unscattered part of the beam as the reference. 

\begin{figure}
\centering
%\adjincludegraphics[width=1\columnwidth,Clip=0bp 0bp 20bp  0bp]{Picture1.png}
\includegraphics[width=0.8\columnwidth]{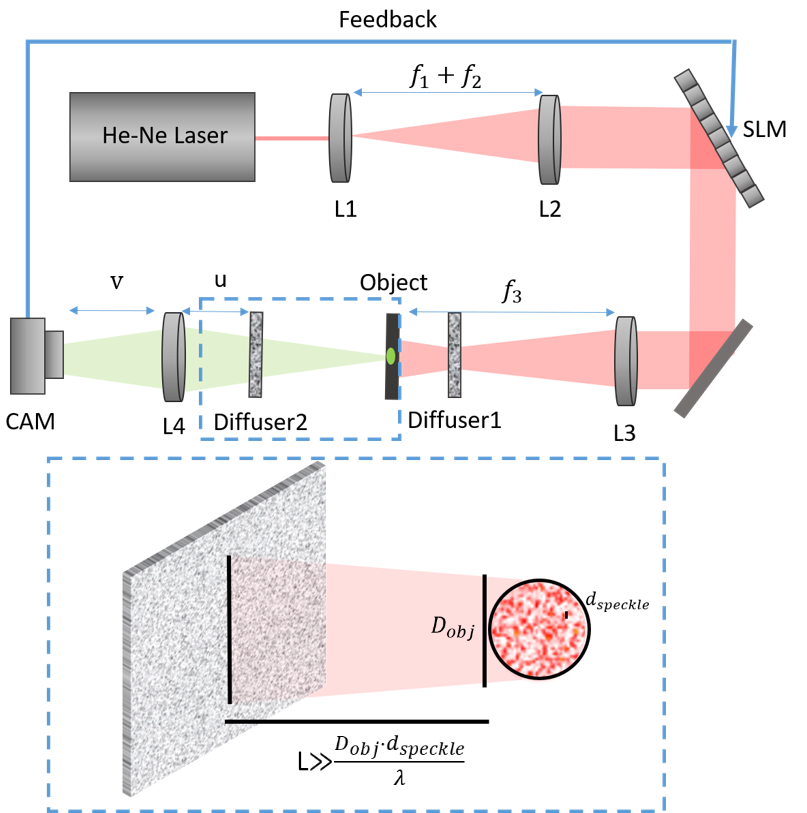} 
\caption{\label{fig:Experimental setup} Setup for focusing between two scattering layers via speckle correlations: a laser beam is shaped by an SLM. The shaped beam passes through a first diffuser, reaches the target object (a transmission-mask), and scattered again by a second diffuser. The second diffuser's surface is imaged on a camera to provide feedback for the focusing process. Inset: close-up view of the considered scenario: a target object with a diameter $D_{obj}$ is illuminated by a speckle field with a  speckle grain size of $d_{speckle}<D_{obj}$. The target is located at a distance of $L$ from the second diffuser. We aim at focusing the light on the target to the diffraction-limit ($d_{speckle}$).}
\end{figure}

To explain the principle of the proposed approach, consider the setup depicted in Fig. \ref{fig:Experimental setup}. Figure 1 presents a simplified scenario where the goal is to use an SLM to focus a monochromatic light beam on a target placed between two highly scattering layers. 
As was recently demonstrated \citep{chang2018single,hofer2018wide}, in the case of a fluorescently labeled target object, estimating the \textit{incoherent} spatial intensity autocorrelation at the object plane is straightforward by following the principles of \cite{incoherent_imaging}. However, using such an approach for iterative focusing is challenging due to the long integration times required to record the low photon flux (per speckle grain) of scattered fluorescence {chang2018single,hofer2018wide}, and due to photo-bleaching. Our approach avoids these limitations, as well as the requirement for fluorescent labeling, by considering \textit{coherent} illumination. Below we show how to estimate the coherent field autocorrelation inside the sample, and how to use it for wavefront-shaping.

%In the following we extend the approach of \cite{incoherent_imaging} to spatially coherent illumination, and show how the estimated coherent field-autocorrelation can be used for iterative focusing.  

In the simplified scenario of Fig. \ref{fig:Experimental setup} a transparent target object placed between two scattering layers is illuminated through the first layer (Diffuser1) by a monochromatic spatially-coherent beam at wavelength $\lambda$. The light intensity at the second layer's external facet is imaged on a camera. Due to the scattering of the first layer, the object is illuminated by a speckle field with complex amplitude $E_{illum}(x,y)$. The field after the object is given by $E_{object}(x,y)=E_{illum}(x,y)T(x,y)$, where $T(x,y)$ represents the thin object's transmission. The light passing through the object propagates a distance $L$ before impinging on the second scattering layer. If $L$ is larger than the far-field distance (an assumption that will be later relaxed), the field illuminating the second layer is proportional to the Fourier transform of the object's field: $E(x',y', L)\propto\mathcal{F}\left( E_{object}(x,y)\right)$.
    %\begin{equation}\label{eq: Fourier}
    %E(x',y', L_1)\propto\tilde{F}\left( E_{object}(x,y)\right)
    %\end{equation}
If the second layer is a thin diffuser that can be modeled as a random phase mask (an assumption that is relaxed below), the light intensity on the far side of the second layer, as imaged by the camera is:

\begin{equation}
\label{eq: cam}
I_{camera}(x',y')\propto \left|E(x',y',L)\right|^2  \propto \left|\mathcal{F}\left( E_{object}(x,y)\right)\right|^2
\end{equation}
i.e. the camera intensity pattern is the power spectrum of the object field. According to the Wiener-Khinchin theorem, the Fourier transform of $I_{camera}(x',y')$ is thus equal to the autocorrelation of the object field, $E_{object}(x,y)$:

\begin{equation}
\label{eq: AC}
 \tilde{F}(I_{camera}(x'',y'')) \propto E_{object}\star E_{object} = A(x,y)  
\end{equation}
Thus the autocorrelation of the field at the object plane, $A(x,y)$, can be directly approximated from a single image of the light intensity on the external facet of the scattering sample (Fig.2). Since the width of the field autocorrelation reflects the width of the illuminated area at the object plane, it can be used as a feedback for iterative focusing via wavefront shaping. The goal of the iterative algorithm is then to 'shrink' the width of the estimated autocorrelation, and thereby produce a sharply focused beam on the object plane (Fig. 3). 

In the above derivation, the second scattering layer was assumed to be infinitely thin and was modeled as a random phase mask. As such, the intensity pattern measured on its external facet is equal to the light intensity impinging on it from the object side, a phenomena known as the "shower-curtain effect" \cite{eitan}.  
In the case of a realistic scattering layer with a finite thickness this simple approximation does not hold. In practice, a point illumination on a diffusive sample of thickness $l$ would result in a speckled blob of diameter $\sim l$ on its output facet \cite{Freund-optica-a,Judkewitz}. Thus, for realistic scattering layers, any features of the input intensity pattern that are finer than $\sim l$ would not be recovered on its external facet. However, features that are coarser than $\sim l$ can be recovered \cite{Freund-optica-a}. Since the camera image, $I_{camera}$, and the estimated autocorrelation are Fourier pairs(Eq. \ref{eq: AC}), the coarse effective resolution of $I_{camera}$ for thick layers limits the autocorrelation estimation to an angular field of view (FoV) no larger than $\theta_{max}\approx\lambda/l$, which is, not by coincidence, the angular range of the memory effect \cite{memory1,memory2Feng,Freund-optica-a}. This limits the target object transverse dimension to ~$L\lambda/l$.

\begin{figure}
\centering
\adjincludegraphics[width=1\columnwidth,Clip=0bp 0bp 0bp  0bp]{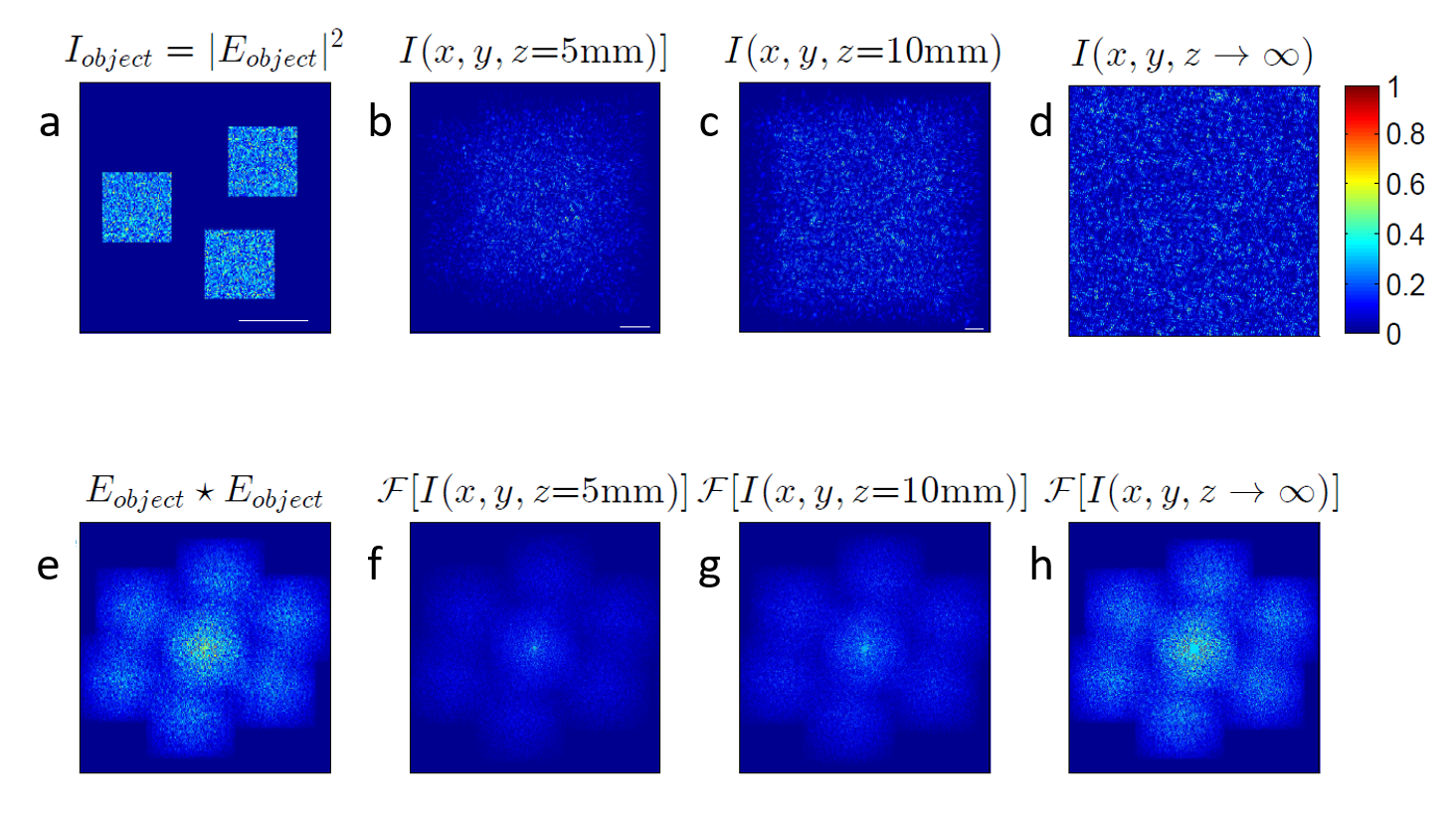}
\caption{\label{fig:concept} Retrieval of a hidden object field autocorrelation from a camera image of the sample surface (numerical example). Top row: evolution of the speckle intensity distribution diffracted from a target hidden object, as measured by a camera: (a) target object plane; (b-d) diffracted lalintensity distribution at increasing distances from the object. Bottom row: comparison between: (e) the object's field autocorrelation, to (f-h) the Fourier transform of the camera images (b-d) (The autocorrelation peak is removed for visualization). The Fourier transform of the speckle intensity provides the object field autocorrelation, and thus a measure of the illuminated area shape. To achieve focusing a wavefront-shaping algorithm minimizes the width of this autocorrelation, concentrating light on the target.}
\end{figure}

In the above simplified derivation $L$ was assumed to be larger than the far-field distance. However, this assumption can be greatly relaxed since the object  is illuminated by a speckle pattern \cite{eitan} or when it is diffusive. Marking the transverse coherence size of the field at the object plane by $d_{speckle}$, where $d_{speckle}$ is either the illumination speckle grain size or alternatively the correlation width of the diffusive object, the diffraction angle of the light propagating from the object is:  $\theta_{diff} \sim sin^{-1}(\frac{\lambda}{d_{speckle}})$. The minimal propagation distance where light from opposite ends of an object mix and interfere to produce the speckles illuminating the second scattering layer is reached at $z_0\approx \frac{D_{obj}}{2\sin(\theta_{diff})} \approx \frac{D_{obj} d_{speckle}}{2\lambda}$, where $D_{obj}$ is the transverse size of the object. For distances larger than $z_0$ ($L>z_0$) each speckle grain in the field impinging on the second scattering layer will result from interferences of all object points, and will have transverse dimensions of  $\sigma_x \approx \lambda L / D_{obj}$. As in the van Cittert-Zernike theorem, the speckle size will then reflect the illuminated object dimensions $D_{obj}$, providing the required feedback for the iterative focusing algorithm. Thus, the autocorrelation of the field at the object plane can be estimated at distances considerably shorter than the far-field distance, as is demonstrated numerically in Fig. \ref{fig:concept}.

To provide proof-of-principle experimental verification of our approach, we constructed the setup of Fig. \ref{fig:Experimental setup}. A Helium-Neon laser beam  (Thorlabs HNL008LB) was expanded and shaped by a phase-only SLM (Holoeye PLUTO) to illuminate a target object (a small transmission mask) placed between two scattering layers (Newport light shaping diffuser, 5° and 10°). The light on the second diffuser external facet was imaged by an sCMOS camera (Andor Zyla 4.2 plus). To inspect the focusing results, a beam-splitter (Thorlabs 45:55 (R:T) Pellicle Beam-splitter) and a second camera (Thorlabs DCU223M, not shown) were used to record the light intensity patterns at the object plane.

\begin{figure}
\centering
\adjincludegraphics[width=1.2\columnwidth,Clip=25bp 180bp 0bp 85bp]{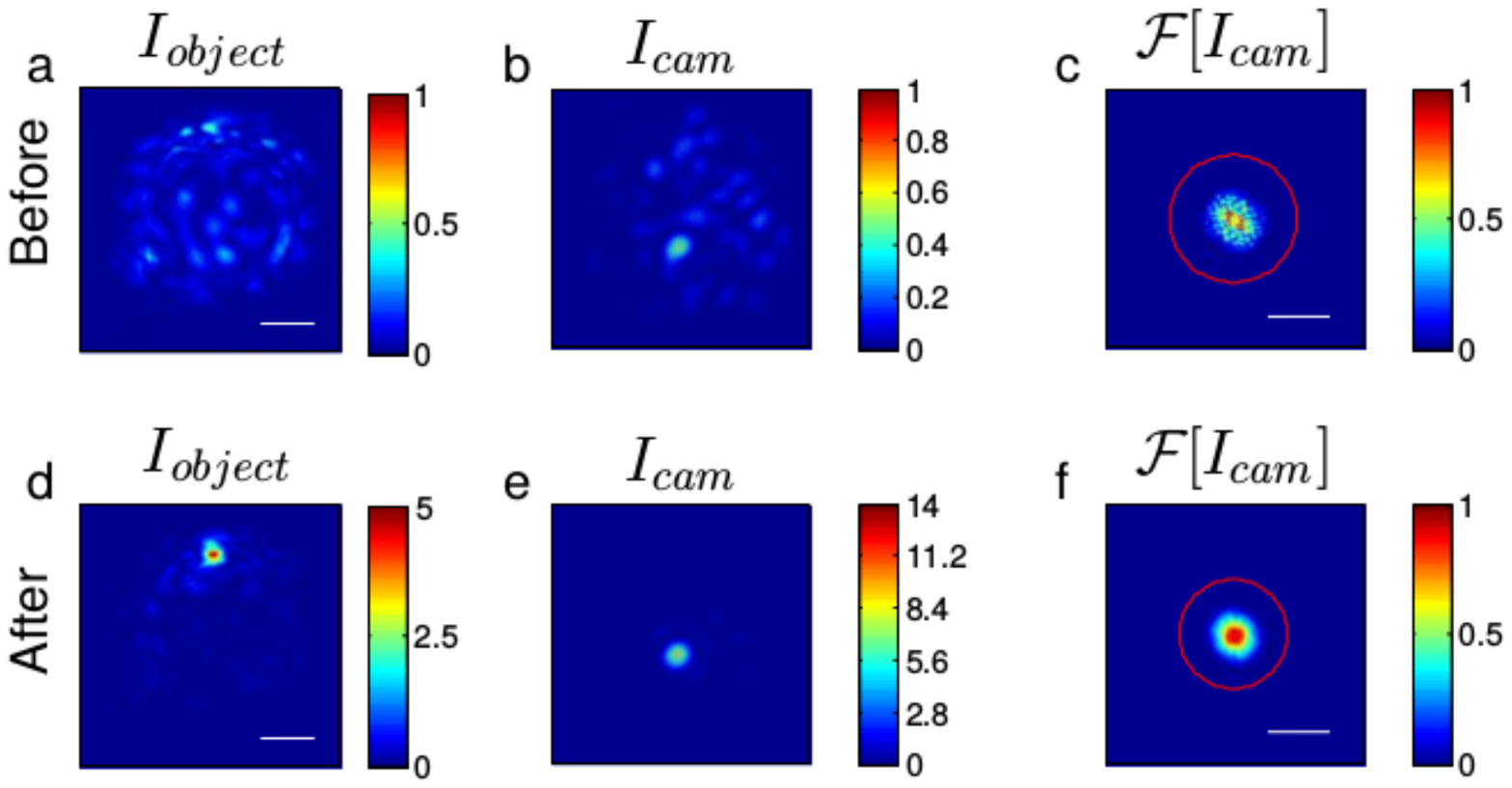}
\caption{\label{fig:Representing set of results 75}
Experimental results with a round target object (\(75 \mu m\) diameter pinhole). Top row: Initial images of the object (a), the camera image (b) and the Fourier Transform of the camera image, which provides an estimate of the object autocorrelation (c). (d-f) same as (a-c) after running the iterative optimization algorithm aimed at minimizing the width of (c) (red circle), resulting in sharp focusing on the object (d). Scalebars, 20 µm.}
\end{figure}

Fig. \ref{fig:Representing set of results 75} displays an experimental result obtained with a circular target object (75  $\mu m$ diameter pinhole). Using the proposed approach, i.e. iteratively shaping the wavefront to shrink the estimated autocorrelation width \ref{fig:Representing set of results 75}(c,f), a sharp focus with a size close to a single speckle grain was obtained \ref{fig:Representing set of results 75}(d), from an initially random speckle pattern (\ref{fig:Representing set of results 75}(a)). The intensity enhancement obtained after ~600 iteration was $\eta\approx 65$.

To shrink the autocorrelation width, we have developed a simple iterative algorithm based on the random partitioning wavefront-shaping algorithm \cite{vellekoop2008phase}: In each iteration of this algorithm, a phase is added to half of the SLM pixels, selected at random. The added phase is cycled in 6 steps from $0$ to $2\pi$. For each phase step, the 'width' of the estimated autocorrelation, $A(r)$ (Eq. \ref{eq: AC}), is approximated by its effective radius: $\langle r\rangle\equiv\sum \frac{A(r)\cdot r}{\sum A(r)}$, after removing the central coherent autocorrelation peak, and cropping the autocorrelation image to a radius proportional to its width, to minimize the background effect. For each iteration, the optimization metric $M =\frac{1}{\langle r \rangle }$, was fitted to a cosine as a function of the added phase, and the optimum phase for the maximal $M$ (i.e. minimal ${\langle r \rangle }$) was kept as the SLM phase pattern. 

%An iterative algorithm was used to minimize the width, as shown in algorithm \ref{alg:algo}.

%\begin{algorithm}
%\caption{Optimization algorithm}\label{alg:algo}
%\begin{algorithmic}[1]
%\Procedure{Phase ramp}{$\Psi_{SLM}, N_{samples}$}
%\State $\frac{N}{2}$ random SLM pixels are selected
%\For $n=1:N_{samples}$  
%\State $\Psi_{SLM}=\Psi_{SLM}+\phi_n$ \Comment $\phi_n=\frac{2\pi n}{N_{samples}}$
%\State Take camera image $I_{cam}$
%\State $AutoCorr=F.T.(I_{cam})$
%\State Calculate $\sigma=\sqrt {\sum\left(\langle r^2\rangle-\langle r\rangle^2\right)}$  
%\State Use $\sigma$ as cutoff radius and remove central peak.
%\State $M=\frac{1}{\langle r \rangle }$ \Comment{Calculated on cut-AutoCorr}
%\State \textbf{fit:} $M(n)\sim \cos (\phi_n-\phi_0)$
%\State \textbf{return} $\phi_0$
%\State \textbf{set} $\Psi_{SLM}=\Psi_{SLM}+\phi_0$
%\EndFor \textbf{end}
%\State \textbf{repeat}
%\EndProcedure
%\end{algorithmic}
%\end{algorithm}

\begin{figure}
\centering
\adjincludegraphics[width=1\columnwidth,Clip=0bp 0bp 0bp  0bp]{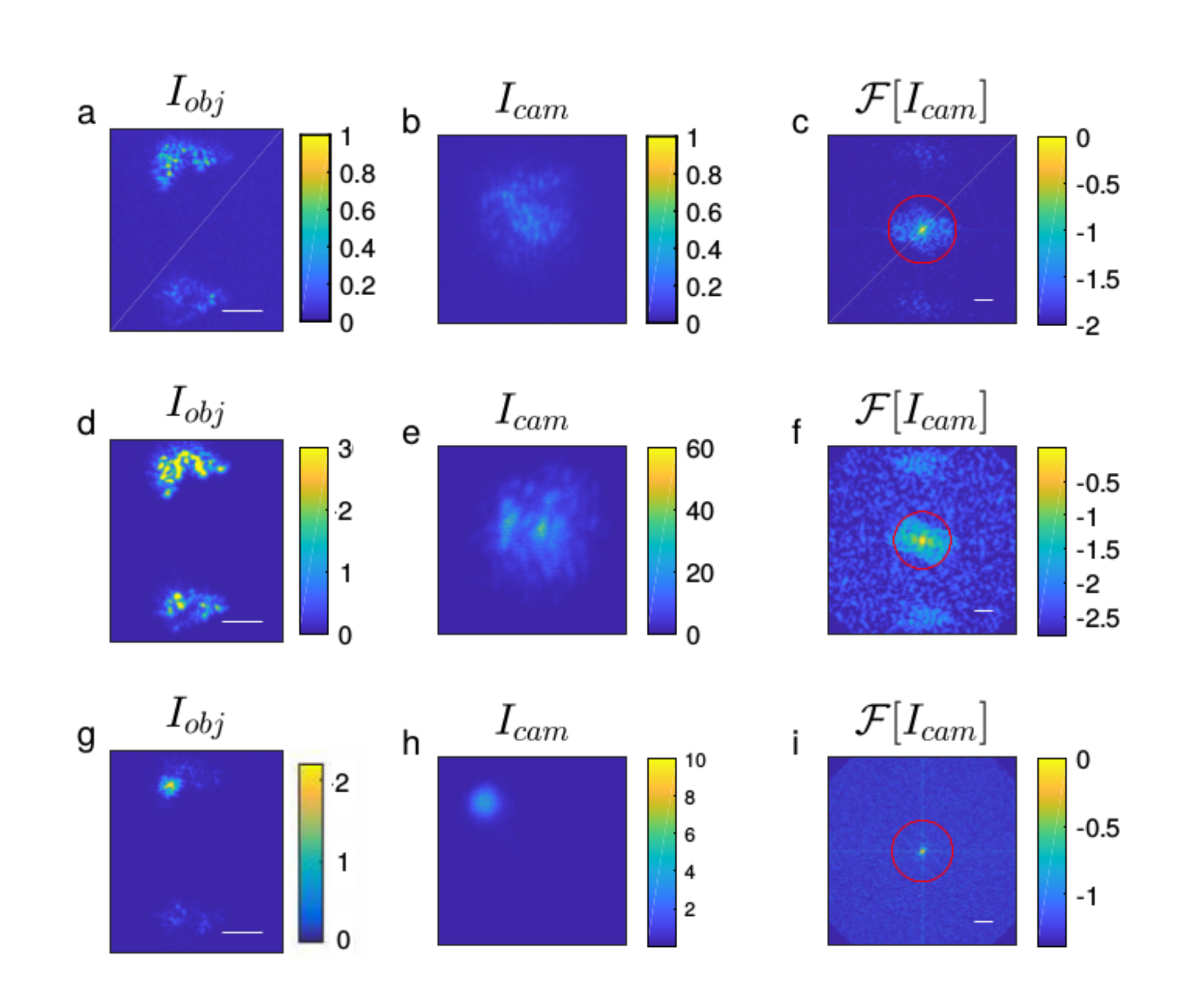}
\caption{\label{fig:sum_vs_opt_holes}Experimental results: Comparison of the proposed approach and conventional optimization of the total intensity on a complex object. Top row: Initial images of the object (a), camera image of the diffuser surface (b), and its Fourier transform (c). (d-f) same as (a-c) after a conventional iterative optimization of the total intensity, (g-i) the same as (d-f) with the proposed approach, showing sharp focusing. 
%(j) evolution of the optimization metrics (red) and the focus enhancement (black), as a function of the iteration step for both approaches. While both metrics are increased, only the proposed approach leads to a corresponding increase in enhancement (\(\eta\)). 
Scalebars: 20 µm.}
\end{figure}

An additional experimental focusing example with a more complex object, and a comparison to conventional optimization is shown in Fig. \ref{fig:sum_vs_opt_holes}. As expected, conventional optimization of the total transmitted intensity (Fig. \ref{fig:sum_vs_opt_holes}(d)) shows enhancement of intensity spread on the entire object area, without forming a localized sharp focus \cite{mosk2}. In contrast, our speckle-correlation based approach (Fig. \ref{fig:sum_vs_opt_holes}(g)) forms a sharp focus. 
%Fig. \ref{fig:sum_vs_opt_holes} (j) compares the evolution of the optimization metrics and the focusing enhancement (\(\eta\))  as a function of the iteration number in the two approaches: while both optimization metrics grow with iterations, only the speckle-correlation guided optimization results in a corresponding increased in focusing enhancement.

Fig. \ref{fig:sum_vs_opt_holes}(c,f,i) display the autocorrelations patterns estimated from the camera images, together with their calculated autocorrelation radius $\langle r\rangle$ (thin red circles). 
%The cropping radius for calculation is $n\cdot\sigma_r=2\sigma_r$ at the beginning, set to include 50\% of the energy. 
Note the two peripheral peaks in Fig. \ref{fig:sum_vs_opt_holes} (c), which reflect the three peaks in the autocorrelation of the of the double-apertured object. 
%A tighter limit would leave these side-lobes out of the autocorrelation width estimation, erroneously reducing the information available for the algorithm, and not allowing it to differentiate between one focus and two foci on both holes. 
In the case of a too small memory-effect range (i.e. too large object), these autocorrelation side-lobes would be impossible to estimate. 

%\textcolor{red}{Suggestion: The strong focus is a result of a long optimization process, as compared to Fig.3, where a plateau was reached much earlier. Since in both cases an invasive focusing will result in higher focus and lower autocorrelation width, we suspect the plateau is an artifact of the algorithm calculation, and a finer tuning of the calculation area will resolve it. In the case of two holes, the calculated $\sigma _r$ is more sensitive to changes due to the autocorrelation shape, thus resolving it by itself.}  

%It can be seen that the metric corresponds well to the contrast measured on the object. The optimization process took ~7000 iterations, The final focus was $PBR=1369$. 
%\textbf{*** HOW COME HERE THE PBR IS >1E3 AND IN THE PREVIOUS OBJECT IT WAS ~60?? THIS IS VERY STRANGE  AND WILL RAISE QUESTIONS***}
%\textcolor{red}{One is 7000 iterations the other is 600. for the round object we reached a plateau quite quickly, maybe because of size, maybe lack of spatial features? The other kept running, so the results are better, but it is understandable}

%It can be seen from fig. \ref{fig:Representing set of results 75}(g)  that the optimization metric corresponds well to the PBR measured on the object, and the optimization process was relatively quick (~600 iterations) for this simple object. The final enhancement is PBR=67.

\begin{figure}
\centering
\includegraphics[width=0.8\columnwidth]{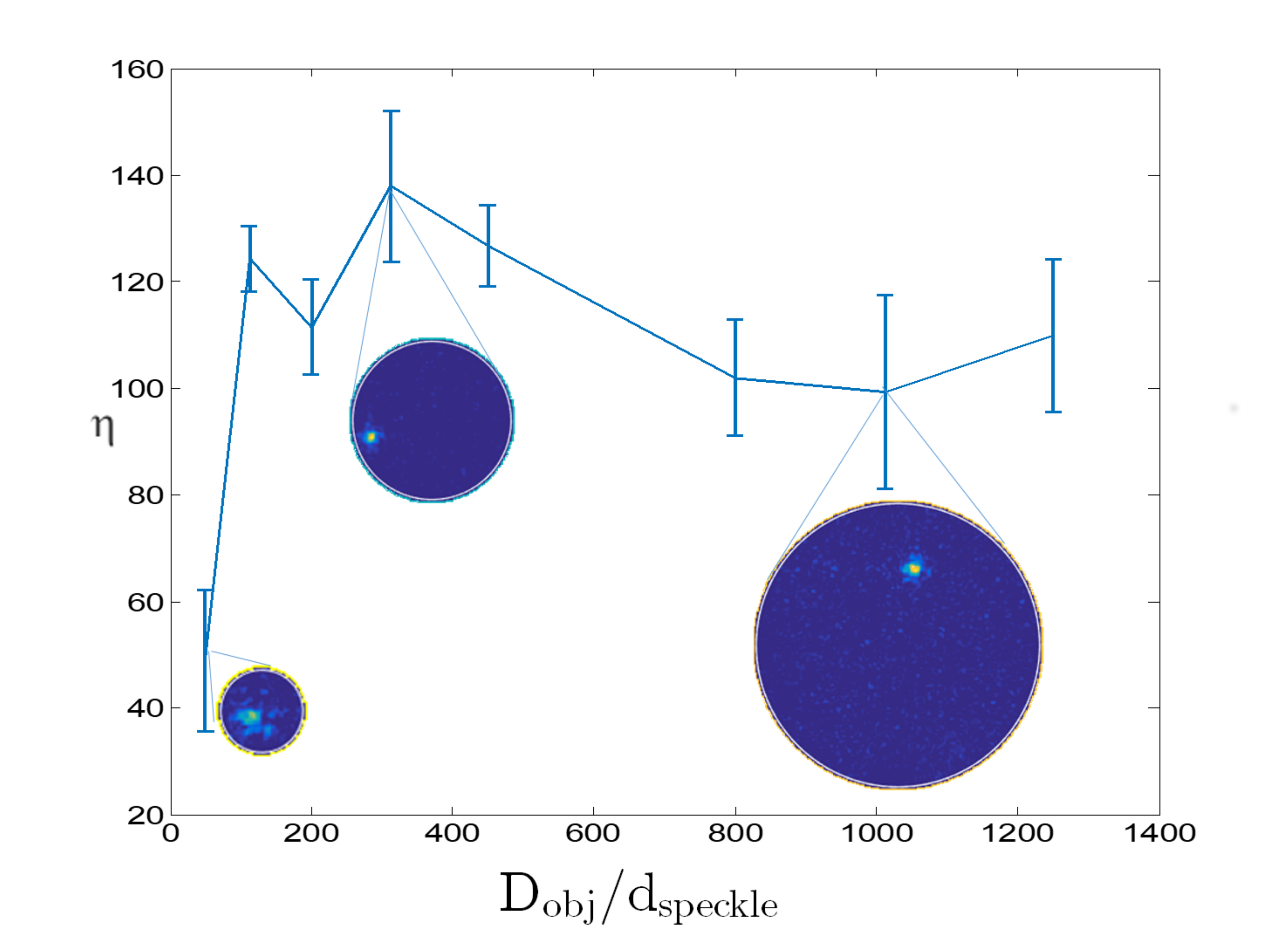}
\caption{\label{fig:size_dif} Numerical comparison between the achieved focus intensity enhancement (\(\eta\)) for different (circular) object sizes. The object size (horizontal axis) is given as the ratio between its area to the speckle grain area, providing an estimate of the number of speckle grains initially contained within the object. Each point is an average of 10 realizations. Error-bars are one standard deviation. Insets: final intensity distribution at the object plane. }
\end{figure}

To study the dependence of the proposed focusing approach on the problem parameters we performed a numerical study of the optimization performance for different object dimensions and the number of controlled SLM pixels. The results of the dependence on the object dimensions are shown in Fig. \ref{fig:size_dif}: each point represents the average enhancement obtained after 3000 iterations, averaged over 10 realizations of the scattering layers. Interestingly, for too small objects, containing only a few speckles grains, the simple proposed focusing algorithm is less effective. We attribute this to be the result of the imperfect removal of the central coherent autocorrelation peak in the  simple analysis we have used: the estimated autocorrelation width of small objects (having dimensions close to the speckle grain size) will be more affected by such simplified peak-removal procedure. However, more robust autocorrelation-width estimation algorithms can be developed for this task. On the other hand, for large objects, we observe only a near constant intensity enhancement (Fig.\ref{fig:size_dif}), and focus dimensions (see Fig.5 insets). Additional simulations with a varying number of controlled SLM pixels, $N_{SLM}$, showed a linear dependence of the enhancement on $N_{SLM}$ (not shown), as in conventional wavefront-shaping with direct imaging of the target.

%\section*{Discussion}
To summarize, we have presented an all-optical guide-star free approach for noninvasive focusing through scattering layers. Unlike previous all-optical noninvasive focusing works \cite{katzOptica,CuiMeng1,F_sharp}, our approach does not require fluorescence labeling or optical nonlinearities. 
%The size of the focus obtained by the presented approach should in principle reach the optical diffraction limit. 

We have demonstrated the approach in proof of principle experiments using thin diffusers. Our proof of principle experiments with thin diffusers should be extendable to volume scattering samples, as long as the field at the target plane is limited to the memory effect FoV. This may be relevant for anisotropically scattering soft tissues at intermediate depths \cite{L_effective}. For objects that are larger than the memory effect range, the information on the longer-distanced parts will not be correctly estimated, and will not allow the simple algorithm to focus to a  single focus.  %and multiple foci separated by a distance larger than the memory effect range. 

%A potential approach to limit the field at the target plane to a spatial extent contained by the memory-effect is via acousto-optic tagging of a small target volume \cite{TRUE}.
%Our approach should in principle allow diffraction-limited focusing inside the large ultrasound tagged area. Such focus could be advantageous for localized probing deep inside scattering media, with increased resolution and signal to noise, breaking the acoustic resolution limit in acousto-optic tagging \cite{TRUE,trove,arxiv}.

%For optimal performance small speckle grains on the object plane will allow a shorter object to scattering layer standoff distance (Fig. \ref{fig:Experimental setup}(b)), and not too large target object dimension, would result in the largest intensity enhancement (PBR) of the focus.

As any iterative optimization approach, the main limitation of the presented approach is the relatively long timescales required for the hundreds of iterations: a typical experiment with the slow refresh rate (~10Hz) liquid crystal SLM used in our experiments required tens of minutes to achieve significant focusing enhancement. However, orders of magnitude faster SLMs \cite{ghielmetti2014direct,lai2015photoacoustically,conkey2012high} and faster cameras would significantly reduce the optimization time. More advanced optimization algorithms, such as genetic algorithms \cite{conkey2012genetic}, may also reduce the number of required iterations.
While sharply-localized foci with dimensions close to the speckle grain were obtained with our simple optimization algorithm, we were not able to experimentally achieve a focus having strictly a single speckle grain dimensions. We attribute this to the relatively slow convergence of the chosen optimization metric and algorithm, and to an imperfect removal of the autocorrelation coherent peak. This is not a fundamental limitation, since we have verified that the autocorrelation of a diffraction limited focus, obtained with direct invasive feedback, indeed posses a smaller width than the optimized foci (not shown). Thus, more advanced optimization metrics are expected to yield improved performances. %\textbf{*** IS THIS LAST SENTENCE TRUE?***}\textcolor{red}{of course it is, our whole paper is based on it}. 
%Thus, the memory-effect provides a novel guide-star for focusing using wavefront-shaping.
%Nonetheless, we were still able to produce an enhancement and received focus intensity of several dozens.
%While we have used transparencies and apertures as the target objects, this was merely an intermediate step, as our long goal would  to use this method in an acousto-optic setup. Where the focused ultrasound wave 'guide-star' serves as a localized deep lying small target. Applying the approach on a selected location defined by the ultrasound guide-star, we expect to be able to increase the peak-intensity and reduce  the size of the formed focus, and thus increase the SNR of related imaging approaches, or find utilization in e.g. photo-dynamic therapy.

\section*{Funding Information}

European Research Council (ERC) Horizon 2020 research and innovation program (grant no. 677909), Azrieli foundation, Human Frontiers Science Program (HFSP).

% Bibliography
\bibliography{paper.bib}

% Full bibliography added automatically for Optics Letters submissions; the following line will simply be ignored if submitting to other journals.
% Note that this extra page will not count against page length
\bibliographyfullrefs{paper}

\end{document}